\documentclass[conference]{IEEEtran}
\IEEEoverridecommandlockouts
\usepackage{hyperref}
\usepackage{cite}
\usepackage{amsmath,amssymb,amsfonts}
\usepackage{algorithmic}
\usepackage{graphicx}
\usepackage{textcomp}
\usepackage{xcolor}
\usepackage{multirow}
\usepackage{comment}
\def\BibTeX{{\rm B\kern-.05em{\sc i\kern-.025em b}\kern-.08em
    T\kern-.1667em\lower.7ex\hbox{E}\kern-.125emX}}
\begin{document}

\title{Acoustic and Perceptual Differences Between\\
Standard and Accented Speech and Their\\
Voice Clones}

\author{
\IEEEauthorblockN{
Tianle Yang$^{1}$, Chengzhe Sun$^{2}$, Phil Rose$^{3}$, Siwei Lyu$^{2}$
}

\IEEEauthorblockA{
$^{1}$Department of Linguistics, University at Buffalo, United States\\
$^{2}$Department of Computer Science and Engineering, University at Buffalo, United States\\
$^{3}$Emeritus Faculty, Australian National University, Australia\\
}
}

\maketitle

\begin{abstract}
Voice cloning is often evaluated in terms of overall quality, but less is known about accent preservation and its perceptual consequences. We compare standard and heavily accented Mandarin speech and their voice clones using a combined computational and perceptual design. Embedding-based analyses showed larger original-clone distances for accented speakers in several speaker-discriminative embedding spaces, but this difference disappeared after adjusting for each speaker's within-original baseline variability. In the perception study, clones are rated as more similar to their originals for standard than for accented speakers, and intelligibility increases from original to clone, with a larger gain for accented speech. These results show that accent variation can shape perceived identity match and intelligibility in voice cloning even when it is not observed in baseline-adjusted speaker-embedding distance, and they motivate treating accent preservation as an explicit component of speaker identity preservation, rather than assuming that it is fully captured by off-the-shelf speaker-discriminative embeddings.
\end{abstract}


\section{Introduction}
Accents\footnote{Linguistically, all varieties have accents; here accented is used in the common sense (regional features relative to the standard variety).} are a natural consequence of dialectal variation. When speech is evaluated against a socially defined standard, systematic phonetic differences and the phonological patterns that organize them are often labeled as ''accent'' \cite{chambers1998dialectology}. In everyday communication, accents can affect intelligibility \cite{munro1995foreign, major2002effects}, social judgments \cite{fuertes2012meta, lev2010don}, and even access to services such as housing \cite{purnell1999perceptual} and employment \cite{hideg2024hear}. In modern speech technology, they also raise a basic but often neglected scientific question: when a system learns a voice, does the source speaker's accent condition affect how voice cloning systems preserve speaker identity and change intelligibility\footnote{Here, intelligibility refers to listeners’ perceived ease of understanding; in parts of the accented-speech literature, this subjective construct is instead termed comprehensibility.}?

This question becomes especially important in voice cloning (audio deepfakes). Voice cloning has clear benefits for speech interfaces, dubbing, accessibility tools, and personalized speech avatars. At the same time, its increasing realism raises security and ethical concerns because cloned voices can be used for impersonation \cite{finley2024deepfake}, fraud \cite{stupp2019fraudsters}, and deception \cite{wang2025one}. Much existing work therefore emphasizes detection and robustness \cite{sun2026survey, nguyen2026analyzing}, speaker modeling and discrimination \cite{YANG2026112768, chen2023voice, chen2025human}, or artifacts introduced by generative models \cite{coletta2025anomaly, salvi2024data, guo2026towards, yang2026assessing}. By contrast, we still have a limited understanding of whether voice-cloning outcomes differ for heavily accented and standard speech, and how such differences are reflected in speaker-identity measures and human perception. This gap matters both for safety (how ''authentic'' a clone is with respect to accent) and for potential applications (whether cloning can improve intelligibility while retaining relative speaker identity for heavily accented speech).

A useful way to approach this problem is to test whether voice cloning alters the relationship between a source speaker and their clone differently for accented and standard speech in speaker-discriminative embedding spaces. Speaker-discriminative embedding systems such as x-vector \cite{snyder2018x}, ECAPA-TDNN \cite{desplanques2020ecapa}, and WavLM \cite{chen2022wavlm} are designed to capture speaker-discriminative information, and they are widely used in speaker verification and related tasks \cite{huang2020speaker}. If voice cloning modifies accent-specific cues in a consistent way, then such modifications might shift how original and cloned utterances relate in embedding space. However, it is not clear how large these shifts are and whether they differ for accented versus standard speech. Quantifying these patterns provides a concrete measurement of embedding-based speaker preservation under different accent conditions.

Similarly, human perception provides a crucial but different window into how voice-cloning technology interacts with speaker identity, accent, and communicative effectiveness. Subjective listening tests have long been central to the evaluation of voice conversion and speech synthesis systems, particularly through listener judgments of naturalness, speaker similarity, and intelligibility \cite{toda2016voice, lorenzo2018voice, sisman2020overview, cooper2024review}. This perceptual dimension has become even more important in the context of audio deepfakes, where listeners do not always reliably distinguish synthetic from genuine speech \cite{mai2023warning}. Recent studies of voice clones further show that cloned voices are not merely technical outputs, but can shape perceived realism, authenticity, familiarity, and social evaluation of the speaker \cite{lavan2025voice, rosi2025perception}.

For accent-related cloning, however, the relevant question is not simply whether a system can normalize an accented voice toward a socially dominant or more widely understood standard. This is because speech perception is not organized around a neutral acoustic space: listeners' linguistic experience shapes how variable acoustic input is mapped onto speech categories \cite{yamada1990perception, levy2009assimilation}, and such biases can support understanding under noise, speaker variation, and other everyday listening conditions \cite{tsao2004speech,kuhl2005early,millet2022self}. Accent is also not simply a matter of regional distance from a standard form; it can index class, social position, community membership, and style \cite{labov1986social,lippi2012english}. A speech pattern that is heard as ''accented'' by standard-accent listeners may be the ordinary and locally appropriate norm for the speaker and their community.

This complexity makes accent-related voice cloning difficult to evaluate through a single notion of ''improvement'' or ''speaker preservation.'' Speaker-discriminative embeddings can show how cloning changes the source-clone relationship in a speaker-discriminative space, but they cannot determine whether listeners perceive the clone as still sounding like the source speaker or as more intelligible. Human perception can test these listener-facing outcomes directly, but it is not an automatic or model-internal measure of how voice cloning reshapes speaker identity. We therefore combine embedding-based analysis with human perception to examine whether accent-conditioned cloning differences appear in computational speaker representations, perceived speaker identity, and intelligibility.

In this paper, we focus on Mandarin speech produced with a heavy regional accent and a socially defined standard accent. Our goal is not to benchmark overall naturalness or synthesis quality, but to characterize whether heavily accented and standard Mandarin speech lead to different voice-cloning outcomes in embedding-based speaker preservation, perceived speaker similarity, and intelligibility. Specifically, we address the following questions:

\begin{enumerate}
  \item \textbf{Accent condition and embedding-based speaker identity:} 
  Does accent condition affect the embedding-space relationship between original speech and its corresponding voice clone, relative to each speaker's own variability in the original recordings?

  \item \textbf{Accent condition and perceived speaker identity:} 
  Does accent condition affect listeners' perceived speaker similarity between original speech and its corresponding voice clone?

  \item \textbf{Accent condition and intelligibility:} 
  Does voice cloning change perceived intelligibility differently for heavily accented and standard Mandarin speech?
\end{enumerate}

By answering these questions, we provide a unified evaluation of how accent condition affects voice-cloning outcomes, linking embedding-based indications of speaker identity with human perception of speaker identity and intelligibility. More broadly, the study tests whether automatic speaker-similarity measures and human judgments of speaker identity remain aligned under accent variation, treating speaker identity preservation as a multi-level construct rather than a single evaluation outcome.

\section{Experimental setting}
In this study, we conduct two experiments: a computational analysis of embedding-based speaker distances, and a perceptual study examining intelligibility and speaker similarity for accented and standard Mandarin speech and their voice clones.

\subsection{Voice-clone generation}
The source materials were drawn from two Mandarin speech corpora. Accented Mandarin recordings were selected from the Mandarin Heavy Accent Speech Corpus \cite{opendatalab}, and standard Mandarin recordings were selected from AISHELL-3 \cite{AISHELL}. From these corpora, we constructed two speaker sets: an accented set consisting of 20 speakers sampled from the Mandarin Heavy Accent Speech Corpus, and a standard set consisting of 20 AISHELL-3 speakers. Because accentedness is difficult to define and directly control as a speaker-level variable, we operationalized the accent condition at the corpus level, contrasting speech from a corpus explicitly described as heavily accented Mandarin \cite{opendatalab} with speech from a standard Mandarin corpus \cite{AISHELL}. For each speaker, we selected 3 evaluation utterances that satisfied the same quality-control criteria: clear speech, no severe background noise or clipping, and sufficient duration for downstream analysis. The standard and accented sets were matched as closely as possible in terms of gender distribution (20 female, 20 male in total), utterance duration, and prompt length.

For each selected evaluation utterance, we generated a cloned version using the corresponding transcript as the synthesis prompt. Here, evaluation utterances refer to the original corpus utterances used as the downstream analysis materials; each evaluation utterance was paired with a cloned counterpart matched in textual content. By contrast, enrollment segments refer to the approximately 20-second audio samples used to condition each voice-cloning system on the target speaker's voice; these segments were separate from the evaluation utterances used for downstream analysis. To reduce nuisance variability across recordings, all enrollment recordings and original evaluation utterances were standardized before they were used in the experiment. Specifically, audio was converted to mono, resampled to a common sampling rate of 16~kHz, and amplitude-normalized. We did not apply additional denoising or manual editing beyond these standardization steps.

We generated cloned speech using three voice-cloning systems: ElevenLabs \cite{elevenlabs_v3}, MiniMax \cite{minimax_voice_cloning}, and AnyVoice \cite{anyvoice_voice_cloning}. For each speaker, the same standardized enrollment segment was used across the three systems, so that system-level differences could not be attributed to differences in enrollment audio. Within each system, synthesis settings were held fixed; when equivalent controls were unavailable across systems, we used the system defaults.

All synthesized outputs were exported and then standardized before downstream analysis using the same mono conversion, resampling, and amplitude-normalization procedure. The same set of original evaluation utterances and cloned counterparts was used in both the embedding-distance analysis and the perception experiment. This design ensures that differences between the computational and perceptual results cannot be attributed to the use of different audio materials across the two analyses.

\subsection{Speaker embedding extraction and distance computation}

We extracted speaker-discriminative embeddings using five pretrained speaker representation models: x-vector \cite{snyder2018x}, ECAPA-TDNN \cite{desplanques2020ecapa}, ResNet-TDNN \cite{villalba2020state}, WavLM-SV \cite{chen2022wavlm}, and UniSpeech-SAT-SV \cite{chen2022unispeech}. These models were selected to test whether the embedding-based results were specific to a single speaker encoder or remained stable across architecturally distinct speaker representation models. They provide a broad, embedding-based proxy for assessing speaker identity preservation, ranging from a conventional TDNN-based x-vector baseline to supervised speaker-verification models such as ECAPA-TDNN and the SpeechBrain ResNet-TDNN model, as well as self-supervised pre-trained models subsequently fine-tuned for speaker verification, including WavLM-SV and UniSpeech-SAT-SV.

For each embedding model, the extracted vectors were L2-normalized before distance computation to remove magnitude effects, ensuring cosine distance primarily reflects angular differences and is comparable within each embedding space. To obtain multiple embedding observations per speaker while keeping comparisons consistent across original and cloned speech, we segment each file into fixed-length tokens using a sliding window (token length $L=3.0$~s; hop size $H=1.5$~s). To reduce the influence of silence, we apply a simple energy-based VAD computed on 30~ms frames with a 10~ms frame shift. A frame is labeled as speech if its root mean square (RMS) energy is greater than $-35$~dB relative to the file-level maximum RMS. We retain a token only if at least 60\% of its frames are labeled as speech. To avoid domination by speakers with many usable tokens, we cap the number of retained tokens at 120 per speaker per condition (original vs.\ cloned) by random subsampling. Overlapping windows from the same utterance were not paired when calculating within-original distances.

For any two embeddings $\mathbf{e}_i$ and $\mathbf{e}_j$, cosine similarity is
\begin{equation}
\cos(\mathbf{e}_i,\mathbf{e}_j)=
\frac{\mathbf{e}_i^{\mathsf{T}} \mathbf{e}_j}{\lVert \mathbf{e}_i\rVert \lVert \mathbf{e}_j\rVert},
\end{equation}
and cosine distance is defined as
\begin{equation}
d_{\cos}(\mathbf{e}_i,\mathbf{e}_j)= 1 - \cos(\mathbf{e}_i,\mathbf{e}_j).
\end{equation}

Within each speaker, we compute two distance sets from available token pairs: original--original distances ($d_{OO}$) among original tokens and original--clone distances ($d_{OC}$) between original and cloned tokens. We summarize each set by its mean (e.g., $\overline{d}_{OC}$) as the primary speaker-level metric. To control computational cost, we use all token pairs when feasible; when the number of pairs is large, we randomly sample up to 25{,}000 within-condition pairs and up to 50{,}000 cross-condition pairs (original--clone) per speaker to estimate the means. Inferential statistics are conducted on speaker-level distances (one aggregated value per speaker per condition), avoiding pseudo-replication from treating token-pair distances as independent observations. Finally, to quantify whether cloning increases within-speaker dispersion in embedding space, we define a speaker-level clone-divergence measure as
\begin{equation}
\Delta_{\mathrm{div}} = \overline{d}_{OC} - \overline{d}_{OO},
\end{equation}
where positive values indicate that original--clone distances exceed the baseline original--original within-speaker distances.

\subsection{Participants}
Participants were recruited for an online perception experiment administered in Qualtrics \cite{qualtrics}. All participants reported being native speakers of Mandarin Chinese who use it as their primary daily language. They also reported normal hearing and no history of speech or hearing disorders. The study protocol was approved by the Institutional Review Board (IRB), and all participants provided informed consent and participated voluntarily. We collected basic demographic information (age group and gender) for descriptive purposes.

A total of $N=138$ participants completed the study. We excluded participants who did not complete the full experiment, reported listening without headphones, or were identified as duplicate submissions. To control for potential accent-familiarity effects in the perception task, we further excluded participants whose self-reported dialect background fell within the same broad Chinese dialect region represented by the accented-speaker stimuli. After all exclusions, $N=67$ participants were retained for analysis (age group: 18--38, $n=43$; 39--59, $n=15$; 60+, $n=9$; gender: female, $n=40$; male, $n=27$).

\subsection{Perception task}
To ensure direct comparability between the computational and perceptual analyses, the perception experiment used exactly the same audio materials as the embedding-distance analysis. Participants completed two rating tasks. For speaker similarity, participants listened to an original utterance and its corresponding clone and rated how similar the clone sounded to the original speaker on a 5-point scale, with higher scores indicating greater perceived similarity. For intelligibility, participants listened to individual original and cloned utterances and rated how easy each utterance was to understand on a 5-point scale, with higher scores indicating greater perceived intelligibility. Trials were presented in randomized order. The similarity ratings were used to evaluate perceived speaker identity preservation, while the intelligibility ratings were used to compute clone-original intelligibility gain for each listener.

\section{Experimental results}
\subsection{Embedding-based speaker distances for accented and standard speech and their voice clones}

Table~\ref{tab:embedding_summary} summarizes the embedding-based analysis across five speaker embedding models. For each embedding model, we computed two speaker-level measures. The first was the original-clone distance, $\bar{d}_{\mathrm{OC}}$, defined as the mean cosine distance between original speech and the corresponding cloned speech. The second was clone divergence, $\Delta_{\mathrm{div}}$, defined as the original--clone distance minus the within-original baseline distance. The $\bar{d}_{\mathrm{OC}}$ measures how far cloned speech is from its source speech in a speaker-discriminative embedding space, whereas $\Delta_{\mathrm{div}}$ measures this distance relative to the amount of variability already present among original utterances from the same speaker.

\begin{table*}[t]
\caption{Embedding-based speaker preservation across five speaker embedding models. Values are estimated marginal means and planned accented--standard contrasts from linear mixed-effects models.}
\label{tab:embedding_summary}
\normalsize
\renewcommand{\arraystretch}{1.15}
\setlength{\tabcolsep}{4.5pt}
\begin{tabular}{lccccc|ccccc}
\hline
\multirow{2}{*}{Embedding model} &
\multicolumn{5}{c|}{Original--clone distance ($\bar{d}_{\mathrm{OC}}$)} &
\multicolumn{5}{c}{Clone divergence ($\Delta_{\mathrm{div}}$)} \\
\cline{2-6} \cline{7-11}
& \textbf{Standard} & \textbf{Accented} & $\Delta_{\mathrm{A-S}}$ & SE & $p_{\mathrm{Holm}}$
& \textbf{Standard} & \textbf{Accented} & $\Delta_{\mathrm{A-S}}$ & SE & $p_{\mathrm{Holm}}$ \\
\hline
ECAPA-TDNN \cite{desplanques2020ecapa} & 0.416 & 0.491 & 0.075 & 0.029 & \textbf{.030} & 0.079 & 0.074 & -0.005 & 0.025 & 1.000 \\
x-vector \cite{snyder2018x} & 0.056 & 0.064 & 0.008 & 0.004 & .118 & 0.010 & 0.015 & 0.005 & 0.004 & 1.000 \\
ResNet-TDNN \cite{villalba2020state} & 0.416 & 0.493 & 0.077 & 0.027 & \textbf{.017} & 0.116 & 0.111 & -0.005 & 0.026 & 1.000 \\
WavLM-SV \cite{chen2022wavlm} & 0.096 & 0.107 & 0.011 & 0.010 & .279 & 0.011 & 0.002 & -0.009 & 0.011 & 1.000 \\
UniSpeech-SAT-SV \cite{chen2022unispeech} & 0.062 & 0.083 & 0.021 & 0.005 & \textbf{$<$.001} & 0.002 & 0.002 & 0.000 & 0.005 & 1.000 \\
\hline
\end{tabular}

\vspace{2pt}
\begin{flushleft}
\footnotesize
Note. Values were estimated from linear mixed-effects models with accent condition, voice-cloning system, and their interaction as fixed effects, and speaker as a random intercept.
$p_{\mathrm{Holm}}$ values are Holm-corrected within each contrast family.
\end{flushleft}
\end{table*}

All statistical tests were performed on speaker-level mean distances, rather than on individual segment-pair distances, to avoid pseudo-replication. We analyzed the embedding-based measures using linear mixed-effects models \cite{bates2015fitting}. For each embedding model and each dependent measure, the model included accent condition (standard vs. accented), voice-cloning system (ElevenLabs, AnyVoice, MiniMax), and their interaction as fixed effects, with speaker included as a random intercept to account for repeated measurements from the same source speaker across cloning systems. Models were fitted separately for each embedding model and each dependent measure, because distance magnitudes are not directly comparable across embedding spaces. The interaction term allowed the accented-standard contrast to vary by voice-cloning system, but the reported contrasts marginalize over systems to address the overall accent-condition effect.

The table reports estimated marginal means for the standard and accented speaker sets, together with the planned accented--standard contrast, its standard error, and the Holm-adjusted $p$-value. Estimated marginal means and contrasts were computed from the fitted mixed-effects model, averaging equally across the three voice-cloning systems. We focus on these planned marginal contrasts rather than on the treatment-coded fixed-effect coefficients, because the primary question was whether accented and standard speakers differed overall in embedding-based speaker preservation, after accounting for voice-cloning system and repeated measurements by speaker. Holm correction was applied separately to the original--clone distance contrasts and the clone-divergence contrasts.

For original--clone distance, all five embedding models showed the same numerical direction: accented speakers had larger $\bar{d}_{\mathrm{OC}}$ values than standard speakers. This accented-standard difference was statistically significant after Holm correction for ECAPA-TDNN ($p_{\mathrm{Holm}}=.030$), ResNet-TDNN ($p_{\mathrm{Holm}}=.017$), and UniSpeech-SAT-SV ($p_{\mathrm{Holm}}<.001$). The same numerical pattern was present for x-vector and WavLM-SV, but these contrasts did not reach statistical significance after Holm correction. Thus, several speaker embedding models indicated larger original-clone distances for accented speakers, although the strength of this pattern varied across embedding architectures.

The clone-divergence results were weaker. After subtracting the within-original baseline distance, accented--standard differences were close to zero across all embedding models, and all Holm-adjusted $p$-values were $1.000$. Thus, although original--clone distances were larger for accented speakers in several embedding spaces, this difference disappeared once baseline within-speaker variability in the original speech was taken into account.

Overall, the embedding-based results suggest a distinction between absolute original-clone distance and baseline-adjusted clone divergence. This pattern suggests that the larger absolute distances for accented speakers may partly reflect greater baseline variability of accented speech in speaker-discriminative embedding spaces, rather than a consistent increase in clone-specific speaker divergence. Once each speaker's own embedding-space variability was taken into account, accented and standard speakers did not differ reliably in clone divergence. These results therefore caution against interpreting off-the-shelf speaker-embedding distances as direct measures of accent preservation. In the five embedding models tested here, the accented–standard difference observed in absolute original–clone distance was no longer present after adjustment for within-original speaker variability. Whether this baseline-adjusted pattern corresponds to listeners’ perception of speaker identity is examined in the following section.

\subsection{Perceptual evaluation of speaker similarity and intelligibility for standard and accented speech and their voice clones}
\begin{figure}[t]
  \centering
  \includegraphics[width=\linewidth]{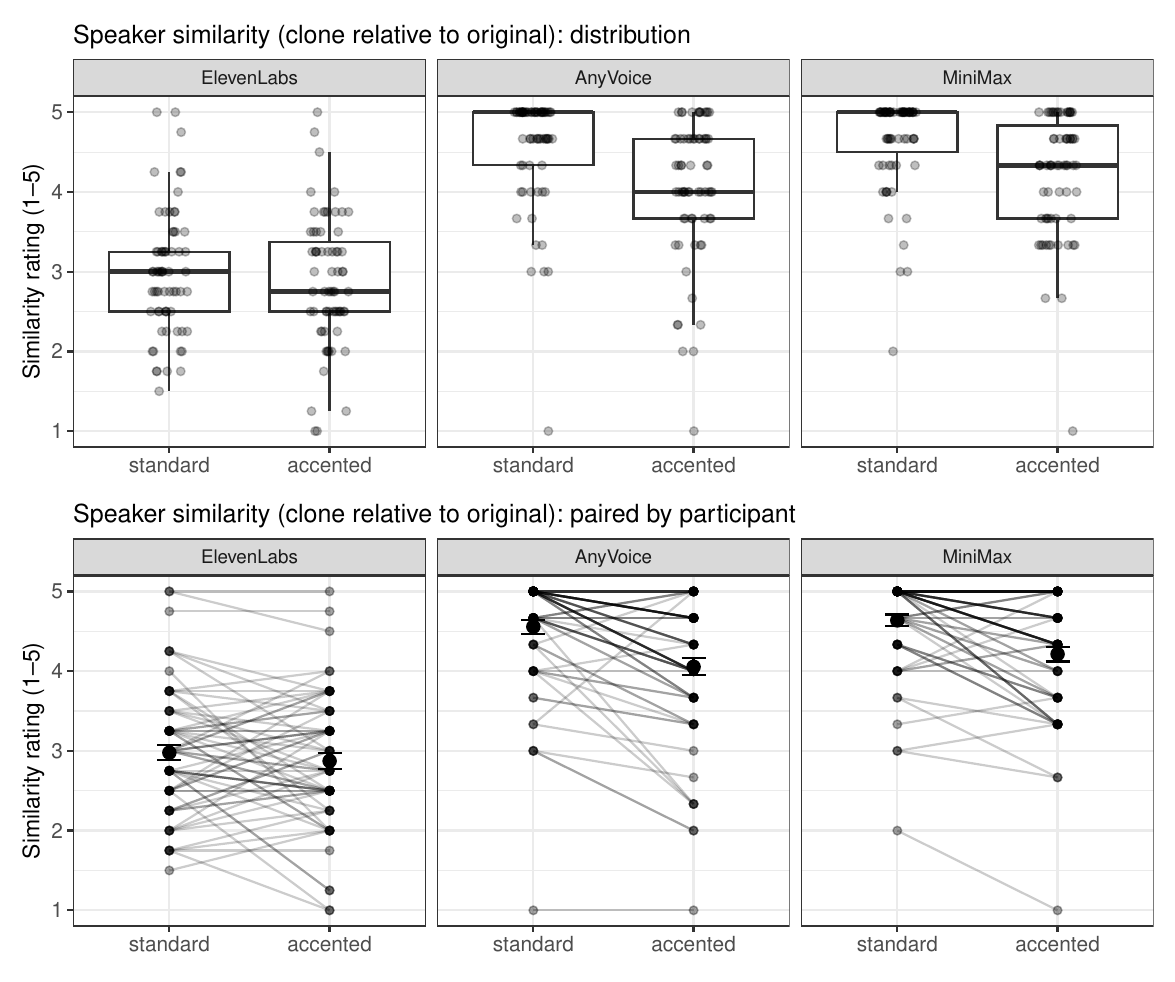}
  \caption{Listener-rated speaker similarity between each voice clone and its corresponding original recording for standard and accented Mandarin speakers, shown separately for different systems. Top: distributions of participant-level mean similarity ratings. Bottom: within-participant paired means for standard vs.\ accented.}
  \label{fig:similarity}
\end{figure}

Figure \ref{fig:similarity} summarizes listeners’ similarity ratings between each voice clone and its corresponding original recording, separately for the standard and accented speaker sets and for each voice-cloning system (ElevenLabs, AnyVoice, MiniMax). The top row shows the distribution of participant-level mean similarity ratings within each condition, while the bottom row shows within-listener changes across the two speaker sets\footnote{Note that values on the plot may fall between integers because each point summarizes a participant’s mean rating across multiple stimuli within the condition.}. 

This analysis directly complements the embedding-distance analysis above. The embedding measures provide a model-internal estimate of how far a clone is from its source speaker in a speaker-discriminative representation, whereas the similarity task tests how far a clone is from its source speaker in human perception. In this sense, the perceptual similarity ratings allow us to ask whether accent-related differences that are weak or absent in off-the-shelf speaker-discriminative embeddings are nevertheless reflected in listener judgments.

Two qualitative patterns stand out. First, similarity ratings are generally higher for AnyVoice and MiniMax than for ElevenLabs, with many ratings near the upper end of the scale in the standard condition, aligning with our earlier embedding-distance analysis, which indicated smaller original–clone divergences for AnyVoice and MiniMax. Second, across systems, perceived similarity between each speaker’s original recording and their corresponding voice clone appears lower for the accented speaker set than for the standard speaker set. Notably, this visual trend does not align with our earlier embedding analysis, which did not show a reliable accent-related change in clone divergence; we therefore test it formally below.

We analysed similarity ratings using a cumulative link mixed model (CLMM; logit link) for the 5-point ordinal scale. The fixed-effects structure included speaker set (standard vs.\ accented), system (ElevenLabs, AnyVoice, MiniMax), and their interaction. The model included random intercepts for listener and stimulus; adding a by-listener random slope for speaker set did not improve fit (AIC$_{\text{RI}}=3132.05$ vs.\ AIC$_{\text{RS}}=3134.65$), so we report the random-intercept model. 

System effects were large: relative to ElevenLabs, similarity ratings were higher for AnyVoice ($\hat{\beta}=3.34$, $z=6.81$, $p<.001$) and MiniMax ($\hat{\beta}=3.51$, $z=7.14$, $p<.001$). Within each system, perceived similarity between each original recording and its corresponding clone was lower for the accented speaker set than for the standard set in AnyVoice (standard$-$accented on the latent logit scale: $\Delta=1.32$, $z=2.54$, $p=.011$) and MiniMax ($\Delta=1.22$, $z=2.35$, $p=.019$), while the same contrast was not reliable for ElevenLabs ($\Delta=0.23$, $z=0.53$, $p=.600$). 

Across systems, the overall standard--accented contrast was also significant when marginalizing over systems (equal-weight averaging: $p=.0008$; proportional-weight averaging: $p=.0018$). In a complementary omnibus likelihood-ratio test, adding the dataset-related terms (main effect plus interactions) significantly improved model fit ($\chi^2(3)=10.23$, $p=.0167$). Overall, listeners rated clones as more similar to their originals for standard speakers than for accented speakers, despite clear differences among voice-cloning systems. This result complements the embedding-distance analysis by showing that accent condition affects perceived speaker identity even when baseline-adjusted speaker-embedding distances do not show a reliable accented-standard difference. In other words, listener judgments appear to be sensitive to accent-related aspects of speaker identity that may not be cleanly represented by off-the-shelf speaker-discriminative embeddings.

For intelligibility, we summarize each listener’s judgments using an intelligibility gain score that compares a clone directly to its matched original. For each participant, system, and speaker set (standard vs.\ accented), we first average the 1--5 intelligibility ratings across items separately for the original and the clone, and then compute gain as $\Delta_{\mathrm{intell}}=\overline{r}_{\mathrm{clone}}-\overline{r}_{\mathrm{orig}}$. Using gain has two advantages: it provides a within-listener, within-item baseline correction that reduces individual differences in rating scale use, and it targets the quantity of interest for voice cloning, namely how much intelligibility changes \emph{relative to the original} rather than absolute ratings.

\begin{figure}[t]
  \centering
  \includegraphics[width=\linewidth]{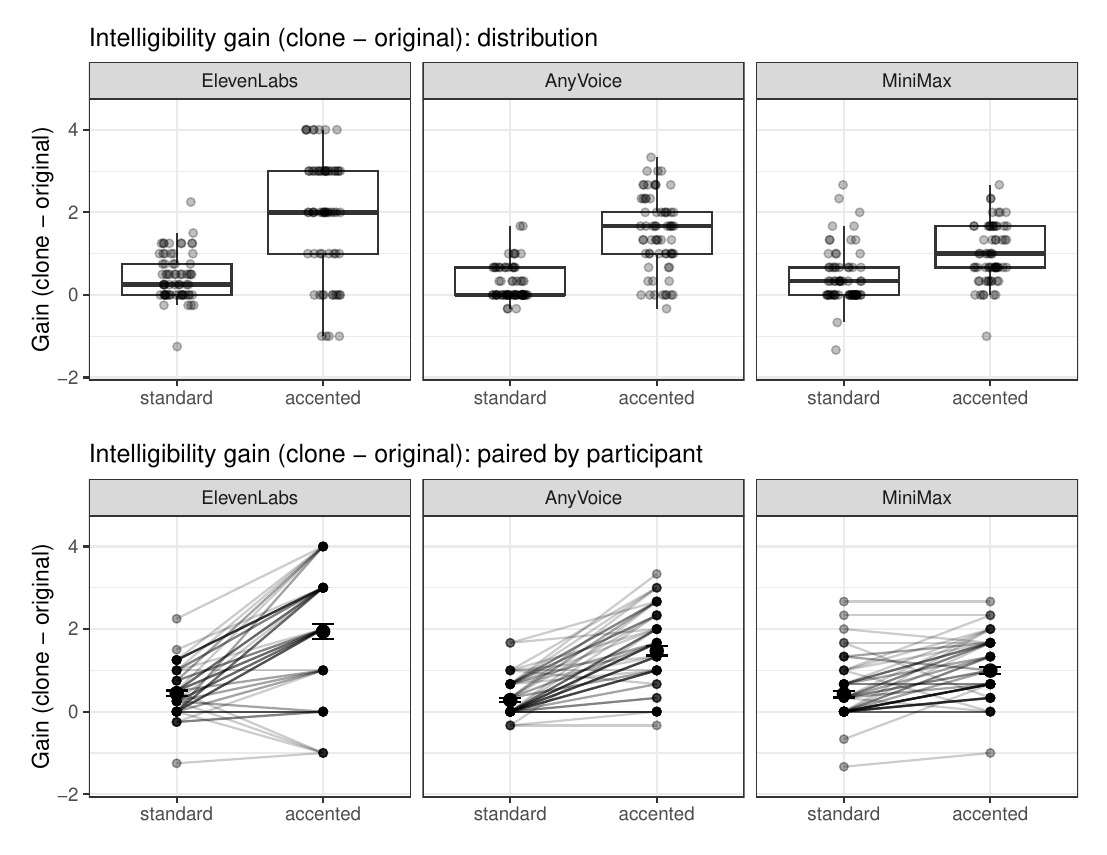}
  \caption{Listener-rated intelligibility gain for each clone relative to its matched original recording, shown separately for speaker sets and for each system. Top: distributions of participant-level mean gains. Bottom: within-participant paired mean gains for Standard vs.\ Accented.}
  \label{fig:intelligibility}
\end{figure}

Figure~\ref{fig:intelligibility} visualizes these gains by system and speaker set. The distributional summaries (top) and within-participant paired means (bottom) indicate that gains are generally positive and tend to be larger for the accented speaker set than for the standard set, with the strength of this pattern varying by system.

We analysed intelligibility ratings using the same CLMM specification as for similarity. Relative to originals, clones received higher intelligibility ratings on the latent logit scale ($\hat{\beta}=2.02$, $z=8.48$, $p<2\times10^{-16}$). Intelligibility ratings were lower for the accented than for the standard speaker set ($\hat{\beta}=-2.12$, $z=-2.77$, $p=.0056$). Importantly, the clone advantage was larger for accented than for standard speech, as indicated by a significant source-by-speaker-set interaction ($\hat{\beta}=1.53$, $z=4.79$, $p=1.65\times10^{-6}$). No system-related effects were reliable (all $p\ge .128$), suggesting that this pattern does not differ clearly across ElevenLabs, AnyVoice, and MiniMax. Overall, clones were rated as more intelligible than their matched originals, and this clone--original advantage was larger for accented than for standard speech.

\section{Discussion}
The results reveal a dissociation between embedding-based source-clone similarity and human perception outcomes in accent-related voice cloning. In the embedding analysis, accented speakers showed larger absolute original-clone distances than standard speakers in several speaker-discriminative embedding spaces, but this difference disappeared after adjusting for each speaker's within-original variability. In human perception, however, listeners rated clones as less similar to their originals for accented speakers than for standard speakers, while also rating cloned speech as more intelligible than original speech, with a larger intelligibility gain for accented speech.

This dissociation suggests that speaker-discriminative embeddings do not capture the full set of cues involved in perceived speaker identity. Embedding distances are useful for measuring whether a clone remains close to its source in a speaker-discriminative representation, but this is not the same as preserving the speaker as heard by listeners. Accent is not simply unwanted variation around a speaker's voice; in speech technologies, it can affect how intelligibility, accentedness, and identity are evaluated, while also contributing to how a speaker is recognized and socially interpreted \cite{huang2023evaluating,lavan2023model,michel2025s}. The absence of an accented–standard difference in baseline-adjusted clone divergence therefore does not mean that accent is irrelevant to cloning or speaker identity, nor should it be taken as evidence that these embedding models contain no accent-related information. Rather, it shows that the tested distance-based evaluation did not reproduce the perceptual effect observed here.

The perceptual results point to a possible identity-intelligibility trade-off. Cloning appears to make accented speech easier to understand, while the same accented clones are also perceived as less similar to their original speakers. This pattern is consistent with an accent-attenuation account: the cloning systems may shift accented speech toward realizations that are more intelligible to the general listener group, while weakening accent-related cues that contribute to perceived speaker identity. This interpretation connects the present study to work on foreign accent conversion, where reducing accentedness while preserving the source speaker's voice has been treated as a central technical goal \cite{zhao2021converting,quamer2022zero}. In voice cloning, however, this goal is not neutral: a clearer clone is not necessarily a more faithful clone, because accentedness, intelligibility, speaker similarity, and perceived identity are separable evaluation dimensions \cite{cooper2024review,wang2024evaluating}.

These findings also have implications for evaluation. Standard voice conversion and speech synthesis evaluations often rely on naturalness and speaker-similarity judgments, while automatic analyses often use speaker-discriminative embeddings or related representation-based measures \cite{zhao2020voice,deja2022automatic,yin2024svsnet}. For accent-related cloning, neither level is sufficient on its own. A system may preserve source-clone proximity in an embedding space while changing how listeners perceive the speaker, or it may improve intelligibility while reducing perceived similarity to the source. Evaluation should therefore distinguish among embedding-based similarity, perceived speaker similarity, and accent preservation, rather than collapsing them into a single quality or preservation score.

\section{Conclusion}
This study shows that accent variation can affect voice-cloning outcomes beyond what is captured by baseline-adjusted speaker-embedding distances. Although accented speakers showed larger absolute original-clone distances in several embedding spaces, these differences disappeared after accounting for each speaker's within-original variability. In contrast, listeners rated accented clones as less similar to their originals, while also rating them as more intelligible than the corresponding original speech. These findings suggest that voice cloning may improve the intelligibility of heavily accented speech, but may do so by altering accent-related speech patterns that contribute to perceived speaker identity. Speaker identity preservation in voice cloning should therefore be treated as a multi-level problem involving computational speaker representations, accent-related speech patterns, and listener judgments, rather than as a single model-internal property or a subjective rating.

\section*{Acknowledgment}
The authors thank all participants in the perception study for their time and contribution to this research.

\newpage
\bibliographystyle{IEEEtran}
\bibliography{mybib}

\end{document}